\documentclass[10pt,letterpaper]{article}
\usepackage{opex3}
\usepackage{amsmath}

\begin{document}

\title{Widely tunable single photon source with high purity at telecom wavelength}

\author{Rui-Bo Jin,$^{1,*}$ Ryosuke Shimizu,$^2$ Kentaro Wakui,$^1$  Hugo Benichi,$^1$  and Masahide Sasaki$^1$}
\address{$^1$National Institute of Information and Communications Technology , 4-2-1 Nukui-Kitamachi, Koganei, Tokyo 184-8795, Japan\\
         $^2$Center for Frontier Science and Engineering, University of Electro-Communications, 1-5-1 Chofugaoka, Chofu, Tokyo 182-8585, Japan}
\email{$^*$ruibo@nict.go.jp}


\begin{abstract}
We theoretically and experimentally investigate the spectral tunability and purity of  photon pairs generated from spontaneous parametric down conversion in periodically poled $\mathrm{KTiOPO_4}$ crystal with group-velocity matching condition.
The numerical simulation predicts that the purity of joint spectral intensity ($P_{JSI}$) and the purity of joint spectral amplitude ($P_{JSA}$) can be kept higher than 0.98 and 0.81, respectively, when the wavelength is tuned from 1460 nm to 1675 nm, which covers the S-, C-, L-, and U-band in telecommunication wavelengths.
We also directly measured the joint spectral intensity at 1565 nm, 1584 nm and 1565 nm, yielding $P_{JSI}$ of 0.989, 0.983 and 0.958, respectively.
Such a photon source is useful for  quantum information and communication systems.
\end{abstract}



\section{Introduction}

As it is well known, classical widely tunable laser sources play  an important role in many fields,  for example, spectroscopy, remote sensing, metrology,  and  optical communications \cite{Duarte2008,Paschotta2008}.
However, the quantum counterpart of the tunable laser, the widely tunable  single photon source, have been less investigated.
Wavelength tunable single photon sources have been generated from quantum dots, but the tunable ranges achieved until now have been narrow \cite{Hoang2012, Benyoucef2009}.

In this paper, we present a widely tunable single photon source based on  spontaneous parametric down conversion (SPDC) in a periodically poled $\mathrm{KTiOPO_4}$ (PPKTP) crystal with group-velocity matching condition.
The concept of group-velocity matching in SPDC was introduced by Keller and Rubin \cite{Keller1997}, and by Grice and Walmsley \cite{Grice1997} in 1997.
PPKTP crystal with group-velocity matching condition was experimentally demonstrated by K{\"o}nig and Wong \cite{Konig2004} for second-harmonic generation  in 2004.
Later, this condition was applied in SPDC for intrinsically pure photon state generation.
Under this condition, the signal and idler photons from SPDC have no spectral correlation, $i.e.$, they are intrinsically pure.
Therefore, there is no need to employ  bandpass filters to obtain high-purity heralded single photons.
Consequently, such photon sources are much brighter than the traditional sources and have showed high brightness in experiments,
as reported by Evans, \emph{et al.}, for entangled photons \cite{Evans2010};  by Eckstein, \emph{et al.}, for two-mode squeezer \cite{Eckstein2011}; and by Yabuno, \emph{et al.}, for four-photon interferometry \cite{Yabuno2012}.

Besides the high purity and high brightness, in this research, we  focus on another important merit of this source: the wide spectral tunability.
From the numerical calculation, we  found  that the  purity of joint spectral intensity ($P_{JSI}$) and the purity of joint spectral amplitude ($P_{JSA}$) can be kept  higher than 0.98 and 0.81, respectively, as the wavelength is tuned from 1460 nm to 1675 nm.
To verify this  simulation, we directly measured the joint spectral intensity,  and obtained high $P_{JSI}$  of  0.989, 0.983 and 0.958 at 1565 nm, 1584 nm and 1565 nm, respectively.

\section{Theory}

\begin{figure}[tbp]
\centering\includegraphics[width=12cm]{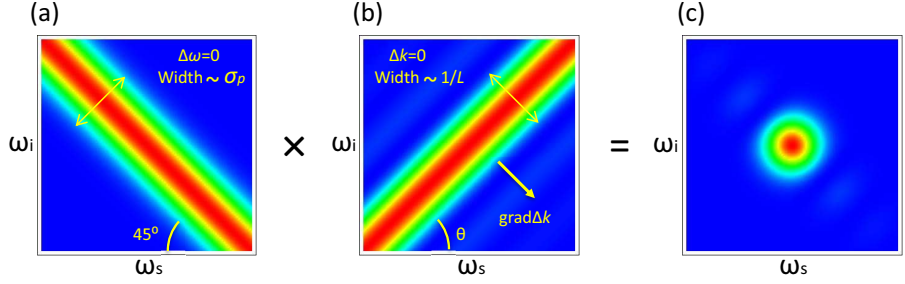}
\caption{ Examples of (a) pump envelope intensity (PEI), (b) phase matching intensity (PMI) and (c) joint spectral intensity (JSI).
(c) is the product of (a) and (b). The width of PEI is proportional to the bandwidth of the pump, while the tilting angle of PEI is fixed, 45 degree to the minus direction of the horizontal axis.
The width of PMI is proportional to the inverse of the crystal length. The direction of PMI is not fixed, but determined by the group velocities of the signal, the idler and the pump. When the width of PEI and PMI are equal and the PMI is at 45 degree to the positive direction of the horizontal axis, JSI can achieve a subcircular shape and maximum purity.
 } \label{JSI}
\end{figure}

In the process of SPDC, one pump photon is split into two lower energy photons, the signal and the idler.
The two-photon component in SPDC can be expressed as \cite{Mosley2008b}
\begin{equation}\label{eq1}
 \left| {\psi _{si} } \right\rangle  = \int\limits_0^\infty  {\int\limits_0^\infty  {d\omega _s } d\omega _i } f(\omega _s ,\omega _i )\hat a_s^\dag  (\omega _s )\hat a_i^\dag  (\omega _i )\left| 0
 \right\rangle,
\end{equation}
where
$ f(\omega _s ,\omega _i )=\phi (\omega _s ,\omega _i)\alpha (\omega _s +\omega _i )$  is the joint spectral amplitude (JSA),  $\phi (\omega _s ,\omega _i)$  and
$ \alpha (\omega _s +\omega _i )$ are the phase matching amplitude and the pump envelope amplitude. $\omega$ is the angular frequency, $\hat a^\dag$ is the creation operator and the subscripts $s$ and $i$ denote the signal and the idler photons, respectively.
Assuming the spectrum of the pump laser has
a Gaussian distribution with a bandwidth of $\sigma_p$, the pump envelope intensity can be written as $|\alpha (\omega _s +\omega _i )|^2 = \exp [ - (\frac{{\omega _s  +\omega _i  - \omega _p }}{\sigma_p })^2 ]$.
Under the collinear condition, the  phase matching intensity can be written in the form of $|\phi (\omega _s ,\omega _i )|^2 = [ \text{sinc}(\frac{{\Delta kL}}{2})]^2 $,
where $\Delta k = k_p  - k_s  - k_i - \frac{{2\pi }}{\Lambda }$ is the difference between the wave vector of the pump ($k_p$), the signal ($k_s$), the idler ($k_i$) and an extra vector ($\frac{{2\pi }}{\Lambda }$) introduced by periodical poling of the crystal.
$L$ and $\Lambda$ are the length and poling period of the SPDC crystal.
Figure \ref{JSI}(a-c)  shows examples of the  pump envelope  intensity $|\alpha (\omega _s +\omega _i )|^2$, phase matching intensity $|\phi (\omega _s ,\omega _i )|^2$ and joint spectral intensity (JSI) $|f(\omega _s ,\omega _i )|^2$.

Considering the Taylor expansion of the three wave vectors
$k_\mu=k_\mu(\omega _{\mu0})+k_\mu ^\prime (\omega _{\mu0})(\omega
_\mu - \omega _{\mu0} )+o((\omega _\mu  - \omega _{\mu0})^2 )$,
($\mu=p, s, i$), the difference of them can be approximated as  $
\Delta k = \Delta k^{(0)}  + \Delta k^{(1)}  +
... $.
In our experiment, the phase matching condition is satisfied at
$\omega_s = \omega_i = \omega_0 = \frac{\omega_p}{2}$ and thus the
zero order contribution disappears:
\begin{equation}\label{eq2}
\Delta k^{(0)}  = k_p (2\omega _0 ) - k_s (\omega _0 ) - k_i
(\omega _0 ) -  \frac{{2\pi }}{\Lambda }= 0 .
\end{equation}
The first order term can be written as
\begin{equation}\label{eq3}
\begin{array}{l}
 \Delta k^{(1)}  = k_p ^\prime (2\omega _0 )(\omega _p  - 2\omega _0 ) - k_s ^\prime (\omega _0 )(\omega _s  - \omega _0 ) - k_i ^\prime (\omega _0 )(\omega _i  - \omega _0 ) \\
  = [k_p ^\prime (2\omega _0 ) - k_s ^\prime (\omega _0 )](\omega _s  - \omega _0 ) + [k_p ^\prime (2\omega _0 ) - k_i ^\prime (\omega _0 )](\omega _i  - \omega _0 ) , \\
 \end{array}
\end{equation}
where ``$\prime$'' represents the first order differentiation with
respect to the angular frequency.

As analyzed in  \cite{Edamatsu2011}, the  slope of the phase
matching condition is determined by the  gradient of $\Delta k$.
\begin{equation}\label{eq4}
\textbf{grad} \Delta k= (\frac{\partial \Delta k}{\partial
\omega_s},\frac{\partial \Delta k}{\partial \omega_i}) =(
V^{-1}_{g,p}(\omega_p)-V^{-1}_{g,s}(\omega_s),
V^{-1}_{g,p}(\omega_p)-V^{-1}_{g,i}(\omega_i) ),
\end{equation}
where the group velocity  $V_{g,\mu}$ is defined as $ V_{g,\mu}  = \frac{{d\omega }}{{dk_\mu(\omega )}} =
\frac{1}{{k_\mu ^\prime (\omega )}}, (\mu=p,s,i).$
From the viewpoint of the tilting angle, $\theta$,
\begin{equation}\label{eq5}
\text{tan}\theta=-\frac{
V^{-1}_{g,p}(\omega_p)-V^{-1}_{g,s}(\omega_s)}{V^{-1}_{g,p}(\omega_p)-V^{-1}_{g,i}(\omega_i)},
\end{equation}
where $\theta$ is the angle between positive direction of horizontal axis
and the ridge direction of the phase matching intensity, as shown in Fig. \ref{JSI}(b).

The purity, a parameter which describes the spectral correlation of the photon source, is defined as $p = Tr(\hat \rho _s^2 ) = Tr(\hat \rho _i^2)$,
where $\hat \rho_s = Tr_i(\left|\psi _{si}\right\rangle\left\langle \psi _{si} \right| )$ is the reduced
density operator of the signal, and ${\rm Tr}_i$ represents the partial trace on the subsystem $i$.
This purity is determined by the factorability of the JSA, $f(\omega _s ,\omega_i )$,
and can be calculated numerically using Schmidt decomposition \cite{Mosley2008b,URen2005}.
By applying Schmidt decomposition on the JSA, we can obtain
\begin{equation}\label{eq6}
f(\omega_1, \omega_2)= \Sigma_j c_j
\phi_j(\omega_1)\varphi_j(\omega_2),
\end{equation}
where  ${ \phi_j(\omega_1)  }$ and ${ \varphi_j(\omega_2)  }$ are two orthogonal basis sets of spectral functions, known as the Schmidt
modes. ${ c_j }$ is a set of real, non-negative, normalized weighting coefficients with $\Sigma_j c_j ^2=1$.
The Schmidt number is defined  as
$
K=\frac{1}{\sum_j c_j ^4 }
$
\cite{Eberly2006}.
$K$ indicates the number of  Schmidt modes existing in the two-photon state and thus can be viewed as an indicator of entanglement.
The purity of the signal or the idler state equals to the inverse of the Schmidt number $K$:
\begin{equation}\label{eq7}
p=P_{JSA}=\sum_j c_j ^4 .
\end{equation}
In experiment, it is difficult to directly measure $f(\omega_1, \omega_2)$, the JSA.
What we usually measured is $|f(\omega _s ,\omega _i )|^2$, the JSI, as have been widely demonstrated in previous experiments \cite{Evans2010, Eckstein2011, Yabuno2012, Mosley2008b}.
To analyze such experimental data in experiment, the Schmidt decomposition can be applied on  $|f(\omega _s ,\omega _i )|^2$,
\begin{equation}\label{eq8}
|f(\omega _1 ,\omega _2 )|^2  = \sum {_j } c _j ^, \phi _j^, (\omega _1  )\varphi _j^, (\omega _2  ).
\end{equation}
We define the purity of the joint spectral intensity as
\begin{equation}\label{eq8}
P_{JSI}  = \sum {_j } c_j^{,4}.
\end{equation}
$P_{JSI}$ is useful in experiment to characterize the joint spectral intensity.
In this paper we focus on  the parameter  $P_{JSI}$.

\section{Numerical simulation}

\begin{figure}[bp]
\centering\includegraphics[width=7cm]{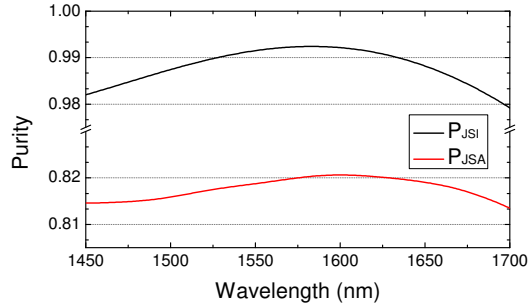}
\caption{ Numerical simulation of the   $P_{JSI}$ and the   $P_{JSA}$ as functions of  the wavelength.  } \label{simu1}
\end{figure}
\begin{figure}[htbp]
\centering\includegraphics[width=12cm]{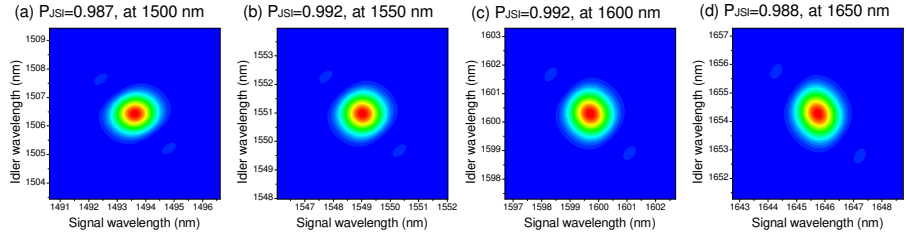}
\caption{ Numerical simulation  of the joint spectral intensity (JSI) at 1500 nm, 1550 nm, 1600 nm and 1650 nm.  } \label{simu2}
\end{figure}

With the Sellmeier equation in \cite{Konig2004}, we calculated  $P_{JSI}$ and $P_{JSA}$  versus  the wavelength, as shown in  Fig. \ref{simu1}.
The horizontal axis is the wavelength of the signal (or the idler, in degenerate case), which is two times of the pump wavelength.
The bandwidth of the pump is set to match the length of the crystal, so as to achieve the maximum purity at 1584 nm.
The optimal bandwidths for  $P_{JSI}$ and $P_{JSA}$  are slightly different.
It is noteworthy that the $P_{JSI}$ and  $P_{JSA}$  can be kept between 0.983 and 0.993, and  0.815 and 0.821, respectively, from 1460 nm to 1675 nm.
The $P_{JSA}$ is smaller than the $P_{JSI}$, mainly caused by the  side lobes in the phase matching amplitude  $\phi (\omega _s ,\omega _i)$.
The  $P_{JSI}$ and  $P_{JSA}$  are relatively low in the regions around 1450 nm and 1700 nm.
This is because  the same bandwidth of the pump is used for all the wavelengths, but this bandwidth is optimum only for 1584 nm.
If the bandwidth of the pump is optimized for 1450 nm or 1700 nm,  we can also achieve a higher values for those regions.
Figure \ref{simu2}(a-d) shows several examples of the  JSI at 1500 nm, 1550 nm, 1600 nm, 1650 nm, respectively.

\section{Experiment and results}

\begin{figure}[bp]
\centering\includegraphics[width=12cm]{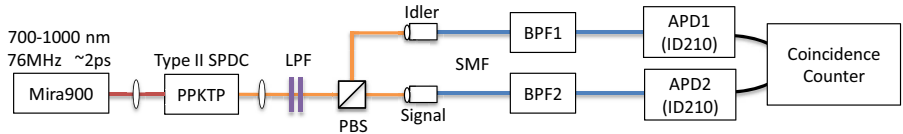}
\caption{ The experimental setup. LPF (long-wave pass filter),  PBS (polarizing beam splitter),
          SMF (single mode fiber),  BPF (bandpass filter),  APD (avalanche photodiodes). } \label{setup}
\end{figure}

The experimental setup is shown in Fig. \ref{setup}.
Picosecond laser pulses (temporal duration $\sim$ 2 ps, central wavelength was tunable from 700 nm to 1000 nm) from a mode-locked Titanium sapphire laser (Coherent, Mira900) were used as the pump source for the SPDC.
Pump pulses with power of 50 mW passed through a 30-mm-long PPKTP crystal with a poling period of 46.1 $\mu$m for type-II SPDC.
The temperature was maintained at $32\,^{\circ}\mathrm{C}$ for the PPKTP crystal.
The down-converted photons, i.e., the signal (vertically polarized) and the idler (horizontally polarized) were separated by a polarizing beam splitter (PBS) and collected into two single-mode fibers (SMF).
Then the photons were filtered by two bandpass filters (BPFs, Optoquest), which had a filter function of Gaussian shape with an FWHM of 0.56 nm  and a tunable central wavelength from 1560 nm to 1620 nm.
Finally, all the collected photons were sent to two InGaAs avalanche photodiode (APD) detectors (ID210, idQuantique) connected to a coincidence counter (Ortec 9353).

To measure the JSI of the photon pairs, we scanned the central wavelength of the BPF1 and BPF2, and recorded the coincidence counts.
BPF1 and BPF2  were  moved in 0.1 nm per step and 60 by 60 steps in all. The coincidence counts were accumulated for 10 seconds for each point.
With the pump wavelength set at 782.5 nm, 792 nm and 807.5 nm, we measured the JSI at 1565 nm,  1584 nm and  1615 nm, respectively.
\begin{figure}[htbp]
\centering\includegraphics[width=12cm]{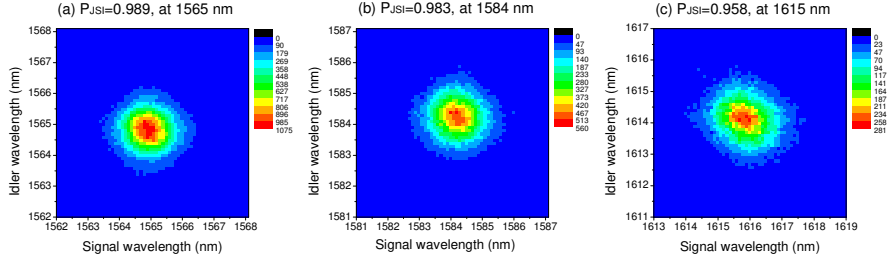}
\caption{ Experimentally measured joint spectral intensity (JSI) at 1565 nm, 1584 nm and 1615 nm.} \label{result}
\end{figure}

The measured $P_{JSI}$ were 0.989, 0.983 and 0.958, respectively, as shown in Fig. \ref{result}.
The maximum coincidence counts rate decreased from 1075 cps at 1565 nm to 281 cps at 1615 nm, due to the decrease of the quantum  efficiency of the APDs.
As depicted in Fig. \ref{simu1}, the theoretical $P_{JSI}$ for 1565 nm,  1584 nm and  1615 nm, were 0.992, 0.992 and 0.991, respectively.
In the simulation of the theoretical $P_{JSI}$,  we assumed that the bandwidth of two BPFs was narrow enough.
However, in experiment the bandwidth of our BPFs (0.56 nm) was comparable to the bandwidth of the signal and idler (1.1 nm).
So, it was necessary to consider  the convolution effect of the two BPFs.
We repeated the simulation with the real filter function and obtained the convolved $P_{JSI}$ as 0.995, 0.995 and 0.995.
The three measured $P_{JSI}$ were very close to these  convolved  values, as shown in Tab. \ref{table1}.
The small disparities may have been caused by the the dark counts in the APDs, or small differences in the BPFs.
\begin{table}[tbp]
\centering\begin{tabular}{c|cccc}
\hline \hline
  &  Theoretical &  Convolved  &  Experimental &  The  \\
  &  $P_{JSI}$    & $P_{JSI}$   & $P_{JSI}$    &   difference   \\
 \hline \\
  1565 nm    &  0.992  &0.995   &  0.989   & 0.006 \\
  1584 nm    &  0.992  &0.995   &  0.983   & 0.012 \\
  1615 nm    &  0.991  &0.995   &  0.958   & 0.037 \\

 \hline \hline
\end{tabular}
\caption{\label{table1} Comparison of the theoretical,  convolved and experimental $P_{JSI}$. }
\end{table}

It was noteworthy to compare the theoretically expected bandwidth and experimentally measured one.
The measured spectrum was a convolution of the original spectrum and the filter function of the BPFs.
The BPFs had a filter function of Gaussian shape with an FWHM of 0.56 nm, while the spectra of the signal and idler were also in a Gaussian shape.
Thus we can calculate the their convolution  as $\text{FWHM}_\text{con}  = \sqrt {0.56^2  + \text{FWHM}_\text{the}^2 }$, where $\text{FWHM}_\text{the}$ was the theoretical calculated  FWHM of the original JSI.
The experimentally measured  $\text{FWHM}_\text{exp}$ can be obtained from the marginal distribution of the data in Fig. \ref{result}.
We compared the theoretical bandwidth and the experimentally measured bandwidth in Tab. \ref{table2}.
The convolved FWHMs were consistent with the measured ones.
The small difference may arise from that the FWHM of pump laser was slightly different from the optimal FWHM in the simulation.
In addition, the  FWHM of the pump laser and  the  FWHM of the two BPFs were slightly wavelength-dependent.
\begin{table}[htbp]
\centering\begin{tabular}{l|cccc}
\hline \hline
  & Theoretical & Convolved   & Experimental  &  The\\
  & $ \text{FWHM}_\text{the} $ &  $\sqrt {0.56^2  + \text{FWHM}_\text{the}^2 }$   & $ \text{FWHM}_\text{exp} $ &   difference \\
 \hline \\
  1565 nm signal & 1.09 nm  &  1.23 nm  &  1.38 nm    & 0.15 nm\\
  1565 nm idler  & 1.13 nm  &  1.26 nm  &  1.32 nm    & 0.06 nm\\
  1584 nm signal & 1.12 nm  &  1.25 nm  &  1.45 nm    & 0.20 nm\\
  1584 nm idler  & 1.12 nm  &  1.25 nm  &  1.40 nm    & 0.15 nm\\
  1615 nm signal & 1.17 nm  &  1.30 nm  &  1.56 nm    & 0.26 nm\\
  1615 nm idler  & 1.10 nm  &  1.23 nm  &  1.43 nm    & 0.20 nm\\
 \hline \hline
\end{tabular}
\caption{\label{table2} Comparison of the convolved bandwidth and the experimentally measured bandwidth. }
\end{table}

\section{Discussion and future}

We need to clarify that purity of joint spectral intensity, $P_{JSI}$, is different from the traditional definition of spectral purity $p$, or $P_{JSA}$.
There is no information on phase of the two-photon wave packet in JSI.
Any phase would reduce the purity of the heralded single photons by introducing temporal correlations.
Thus, we cannot directly obtain  purity of  the heralded single photons from JSI.
As shown in Fig. \ref{simu1}, $P_{JSA}$ is usually smaller than $P_{JSI}$, due to the side lobes in the JSA.
However, the detrimental effect of the side lobes is not severe, according to the previous experiments.
Several groups have reported  high-visibility  interference between the signal and idler photons from such PPKTP crystals, e.g., visibility has achieved $94\%$ in \cite{Evans2010} and  $98\%$ in \cite{Yabuno2012}.
Especially,  the interference visibility was  $95\%$ for both PPKTP crystal and custom-poled crystal in \cite{Branczyk2011}.

To reduce the effect of side lobes in JSA, a new method of engineering the nonlinearity profile of the poled crystals has been proposed recently \cite{Branczyk2011}.
With this method, the phase matching amplitude can be tailored to approximate a Gaussian function.
Therefore, the side lobes can be significantly suppressed and the $P_{JSA}$  can be improved from 0.81 to 0.99 at 1576 nm \cite{Branczyk2011}.
They used higher-order poling to modulate the crystal. In this process, the material property of KTP was not changed.
Therefore, in principle, widely tunable photon sources with higher purity could   be generated with such custom-poled crystal.
However, this method had the drawbacks of smaller effective nonlinearity and thus relatively lower photon pair rate \cite{Branczyk2011} compared to  the traditional PPKTP crystals.
Alternatively, the purity can be improved by using a wide-band BPF to filter out the side lobes, but this method slightly decreases the count rate.

This spectrally pure photon source can be easily transformed into an entangled photon source by adding two calcite prisms as displacers \cite{ Evans2010, Fiorentino2008} or setting it into a Sagnac loop configuration \cite{Fedrizzi2007, Predojevic2012}.
This photon source can be used for  pulsed two-mode squeezed state generation by combining the signal and idler in the degenerate condition \cite{Gerrits2011}.
Another possible application for this photon source is wavelength-multiplexing based multiparty quantum communication system.
If this PPKTP crystal is pumped by a laser frequency comb, e.g., with wavelengths distributed from 730 nm to 835 nm, we can generate  serial  pure photons distributed from 1460 nm to 1675 nm.
These photons can be used for a wavelength-multiplexed system, which is one step for  practical multiparty quantum communication systems.
The classical broadly tunable laser source have showed a tremendous impact in many and diverse fields of science and technology \cite{Duarte2008}.
As the quantum counterpart of classical broadly tunable laser source, our widely  tunable single photon source also has the potential to play an important role in quantum information and communication systems, when the wavelength tunability are necessary.

\section{Summary}

In summary, we have demonstrated the generation of widely tunable and highly pure  photons from PPKTP crystal.
In the theoretical simulation, we found the wavelength can be tuned from 1460 nm to 1675 nm, with $P_{JSI}$ over 0.98 and $P_{JSA}$ over 0.81.
In experiment, we measured the JSI at 1565 nm, 1584 nm and 1565 nm, and achieved $P_{JSI}$ of 0.989, 0.983 and 0.958, respectively.
This result was  well consistent with our theoretical simulation.
We also discussed the future application of this source.

\section*{Acknowledgments}
The authors are grateful to M. Yabuno, P. G. Evans and T. Gerrits for helpful discussions.
This work was supported by the Founding Program for World-Leading Innovative R\&D on Science and Technology (FIRST).


\begin{thebibliography}{99}
\newcommand{\enquote}[1]{``#1''}

\bibitem{Duarte2008}
F.~J. Duarte, ed., \emph{Tunable Laser Applications} (CRC Press, 2008), 2nd ed.

\bibitem{Paschotta2008}
R.~Paschotta, \emph{Encyclopedia of Laser Physics and Technology} (Wiley-VCH,
  2008).

\bibitem{Hoang2012}
T.~B. Hoang, J.~Beetz, M.~Lermer, L.~Midolo, M.~Kamp, S.~H{\"o}ling, and
  A.~Fiore, \enquote{Widely tunable, efficient on-chip single photon sources at
  telecommunication wavelengths,} Opt. Express \textbf{20}, 21758--21765
  (2012).

\bibitem{Benyoucef2009}
M.~Benyoucef, H.~S. Lee, J.~Gabel, T.~W. Kim, H.~L. Park, A.~Rastelli, and
  O.~G. Schmidt, \enquote{Wavelength tunable triggered single-photon source
  from a single cdte quantum dot on silicon substrate,} Nano Letters
  \textbf{9}, 304--307 (2009).

\bibitem{Keller1997}
T.~E. Keller and M.~H. Rubin, \enquote{Theory of two-photon entanglement for
  spontaneous parametric down-conversion driven by a narrow pump pulse,} Phys.
  Rev. A \textbf{56}, 1534--1541 (1997).

\bibitem{Grice1997}
W.~P. Grice and I.~A. Walmsley, \enquote{Spectral information and
  distinguishability in type-II down-conversion with a broadband pump,} Phys.
  Rev. A \textbf{56}, 1627--1634 (1997).

\bibitem{Konig2004}
F.~K{\"o}nig and F.~N.~C. Wong, \enquote{Extended phase matching of
  second-harmonic generation in periodically poled $\mathrm{KTiOPO_4}$ with
  zero group-velocity mismatch,} Appl. Phys. Lett. \textbf{84}, 1644--1646
  (2004).

\bibitem{Evans2010}
P.~G. Evans, R.~S. Bennink, W.~P. Grice, T.~S. Humble, and J.~Schaake,
  \enquote{Bright source of spectrally uncorrelated polarization-entangled
  photons with nearly single-mode emission,} Phys. Rev. Lett. \textbf{105},
  253601 (2010).

\bibitem{Eckstein2011}
A.~Eckstein, A.~Christ, P.~J. Mosley, and C.~Silberhorn, \enquote{Highly
  efficient single-pass source of pulsed single-mode twin beams of light,}
  Phys. Rev. Lett. \textbf{106}, 013603 (2011).

\bibitem{Yabuno2012}
M.~Yabuno, R.~Shimizu, Y.~Mitsumori, H.~Kosaka, and K.~Edamatsu,
  \enquote{Four-photon quantum interferometry at a telecom wavelength,} Phys.
  Rev. A \textbf{86}, 010302 (2012).

\bibitem{Mosley2008b}
P.~J. Mosley, J.~S. Lundeen, B.~J. Smith, and I.~A. Walmsley,
  \enquote{Conditional preparation of single photons using parametric
  downconversion: a recipe for purity,} New Journal of Physics \textbf{10},
  093011 (2008).

\bibitem{Edamatsu2011}
K.~Edamatsu, R.~Shimizu, W.~Ueno, R.-B. Jin, F.~Kaneda, M.~Yabuno, H.~Suzuki,
  S.~Nagano, A.~Syouji, and K.~Suizu, \enquote{Photon pair sources with
  controlled frequency correlation,} Progress in Informatics \textbf{8}, 19--26
  (2011).

\bibitem{URen2005}
A.~B. U'Ren, C.~Silberhorn, K.~Banaszek, I.~A. Walmsley, R.~Erdmann, W.~P.
  Grice, and M.~G. Raymer, \enquote{Generation of pure-state single-photon
  wavepackets by conditional preparation based on spontaneous parametric
  downconversion,} Laser Physics \textbf{15}, 146--161 (2005).

\bibitem{Eberly2006}
J.~Eberly, \enquote{Schmidt analysis of pure-state entanglement,} Laser Physics
  \textbf{16}, 921--926 (2006).

\bibitem{Branczyk2011}
A.~M. Bra\'{n}czyk, A.~Fedrizzi, T.~M. Stace, T.~C. Ralph, and A.~G. White,
  \enquote{Engineered optical nonlinearity for quantum light sources,} Opt.
  Express \textbf{19}, 55--65 (2011).

\bibitem{Fiorentino2008}
M.~Fiorentino and R.~G. Beausoleil, \enquote{Compact sources of
  polarization-entangled photons,} Opt. Express \textbf{16}, 20149--20156
  (2008).

\bibitem{Fedrizzi2007}
A.~Fedrizzi, T.~Herbst, A.~Poppe, T.~Jennewein, and A.~Zeilinger, \enquote{A
  wavelength-tunable fiber-coupled source of narrowband entangled photons,}
  Opt. Express \textbf{15}, 15377--15386 (2007).

\bibitem{Predojevic2012}
A.~Predojevi\'{c}, S.~Grabher, and G.~Weihs, \enquote{Pulsed sagnac source of
  polarization entangled photon pairs,} Opt. Express \textbf{20}, 25022--25029
  (2012).

\bibitem{Gerrits2011}
T.~Gerrits, M.~J. Stevens, B.~Baek, B.~Calkins, A.~Lita, S.~Glancy, E.~Knill,
  S.~W. Nam, R.~P. Mirin, R.~H. Hadfield, R.~S. Bennink, W.~P. Grice,
  S.~Dorenbos, T.~Zijlstra, T.~Klapwijk, and V.~Zwiller, \enquote{Generation of
  degenerate, factorizable, pulsed squeezed light at telecom wavelengths,} Opt.
  Express \textbf{19}, 24434--24447 (2011).

\end{thebibliography}
\end{document}